\documentclass[twocolumn,aps,epsfig,nofootinbib]{revtex4}

\usepackage{graphicx}
\usepackage{epstopdf}
\usepackage{latexsym}
\usepackage{amssymb}
\usepackage{amsmath}
\usepackage{color}
\usepackage{mathrsfs}
\usepackage[center]{subfigure}

\begin{document}

\newcommand{\bq}{\begin{equation}}
\newcommand{\eq}{\end{equation}}
\newcommand{\bqn}{\begin{eqnarray}}
\newcommand{\eqn}{\end{eqnarray}}
\newcommand{\nb}{\nonumber}
\newcommand{\lb}{\label}
\newcommand{\PRL}{Phys. Rev. Lett.}
\newcommand{\PL}{Phys. Lett.}
\newcommand{\PR}{Phys. Rev.}
\newcommand{\PRD}{Phys. Rev. D}
\newcommand{\CQG}{Class. Quantum Grav.}
\newcommand{\JCAP}{J. Cosmol. Astropart. Phys.}
\newcommand{\JHEP}{J. High. Energy. Phys.}
\newcommand{\PLB}{Phys. Lett. B}
 
\title{Observable acceleration of jets by a Kerr black hole}

\author{J. Gariel$^1$}
\email{jerome.gariel@obspm.fr}

\author{ N. O. Santos $^{1,3}$}
\email{n.o.santos@qmul.ac.uk}

\author{Anzhong Wang $^{4, 5, 6}$}
\email{anzhong_wang@baylor.edu}

\affiliation{ $^{1}$Sorbonne Universit\'es, UPMC Universit\'e Paris 06, LERMA, UMRS8112 du CNRS, Observatoire de Paris-Meudon 5, 
Place Jules Janssen, F-92195 Meudon Cedex, France\\
$^{3}$School of Mathematical Sciences, Queen Mary, University of London, London E1 4NS, UK\\
$^{4}$GCAP-CASPER, Department of Physics, Baylor University, Waco, Texas 76798-7316, USA\\
$^{5}$ Institute  for Advanced Physics $\&$ Mathematics, Zhejiang University of Technology,
Hangzhou 310032,  China\\
$^{6}$ Departamento de F\'{\i}sica Te\'orica, Instituto de F\'{\i}sica, UERJ, 20550-900, Rio de Janeiro, Brazil}

\begin{abstract}

In the framework of a model based on the gravitational field of the Kerr black hole, we turn to investigate the kinematic behavior of extragalactic jets. 
We analytically calculate the observable velocities and accelerations along any geodesic. Then, by numerical calculations, we apply our results to
a geodesic, typical of the M87 jet, and probe our results by confrontation to recent observations.  A transition from non-relativistic to ultrarelativistic 
speeds at subparsec scale is highlighted. This transition comes sooner and more abruptly than in models based on magnetic paradigm, which 
indicates that we need a weaker magnetic field to explain observed synchrotron radiation. We attribute the ejection phenomenon to the repulsive
effect of the gravitomagnetic Kerr field.

\end{abstract}

\maketitle

\section{Introduction}

The extragalactic jets are important astrophysical phenomena generated by central engines supposed to be black holes (BH). Inspired by the theory of jets ejected from 
less exotic sources, as usual stars and neutron stars, the first models were based on the electromagnetic field, with magnetic field lines anchored in the BH horizon \cite{Blandford}, 
or in the accretion disk \cite{Blandford1}, including magnetohydrodynamics \cite{Punsly}. Until today, these standard models remain the main paradigm used as basis for 
confrontations with the observations. The raised questions for probing the models concern, inter alia, the shape, the length and permanence, the profile, the power, the radiation, 
the composition, the acceleration, and the birth of the jet. Here, we shall focus on the acceleration at the launching.

The initial acceleration of the jet is currently assumed to be of magnetic origin, and all the observations are confronted to models inside this framework \cite{Lee,Asada}.
We propose a model mainly based on the gravitational field. In General Relativity (GR) the only axisymmetric stationary metric with a good asymptotic behavior is the Kerr BH. 
We first analyze  the admissible collimation \cite{Gariel}, the high energies \cite{Pacheco}, the profile of the M87 jet \cite{Barba}, and the Penrose effect \cite{Gariel1}. Then,
 we turn to kinematical studies by considering the proper acceleration along the axis of symmetry, showing the clear existence of a repulsive effect \cite{Gariel2}. Here we deepen 
 this study by calculating the observable velocities and accelerations. After recalling the basic equations (section II) and previous results (section III), we analytically calculate the 
 proper acceleration along the radial $\rho$-component (section IV) and the temporal variation (Section V), the observable velocity and acceleration (Section VI). Then, by numerical 
 calculations, we apply our results to a geodesic, typical of the M87 jet as studied in \cite{Barba}, and compare  our results with the observations \cite{Lee,Asada} (Section VII). 
In Section VIII, we summarize our main results with a brief discussion.


\section{Recalling the Kerr geodesics in Weyl coordinates}

The Kerr metric given in the usual Boyer-Lindquist spherical coordinates $\bar{r}$, $\theta $ and $\phi $ reads
\begin{eqnarray}
ds^2=-\left( \frac{{\bar{r}}^2-2M{\bar{r}}+a^2}{{\bar{r}}
^2+a^2\cos^2\theta }\right) (d{\bar t}-a\sin ^2\theta \,d\phi )^2 \nonumber\\
+\frac{\sin ^2\theta }{{\bar{r}}^2+a^2\cos \theta ^2}\left[ a\,d{\bar t}-({\bar{r}}^2+a^2)\,d\phi \right]^2 \nonumber\\
+({\bar{r}}^2+a^2\cos ^2\theta )\left( \frac{d{\bar{r}}^2}{{\bar{r}}^2-2M{\bar{r}}+a^2}+d\theta^2\right),  \label{1}
\end{eqnarray}
where $M$ and $J \equiv Ma$ are, respectively, the mass and the angular momentum of
the source, and we have taken units such that $c=G=1$ where $G$ is Newton's
constant of gravitation. Rescaling the $\bar t$ and $\bar{r}$ coordinates as $t={\bar t}/M$ ($s={\bar\tau}=M\tau$) and $r={\bar{r}}/M$ the timelike geodesics equations are
\begin{eqnarray}
{\dot{r}}^2&=&(a_4r^{4}+a_3r^3+a_2r^2+a_1r+a_0)\nb\\
&& \times
\left[r^2+\left(\frac{a}{M}\right)^2\cos^2\theta\right]^{-2},  \label{2}
\\
{\dot{\theta}}^2&=&\frac{b_4\cos^4\theta+b_2\cos^2\theta+b_0}{1-\cos^2\theta }\nb\\
&& \times \left[r^2
+\left(\frac{a}{M}\right)^2\cos^2\theta\right]^{-2},  \label{3}
\end{eqnarray}
with coefficients
\begin{eqnarray}
a_0=-\frac{a^2Q}{M^4}=-\left(\frac{a}{M}\right)^2b_0, \label{4} \\
a_1=\frac{2}{M^2}\left[(aE-L_z)^2+Q\right] ,  \label{5} \\
a_2=\frac{1}{M^2}\left[a^2(E^2-1)-L_z^2-Q\right] ,  \label{6}\\
a_3=2,  \label{7} \\
a_4=E^2-1,  \label{8}
\end{eqnarray}
and
\begin{eqnarray}
b_0=\frac{Q}{M^2},  \label{9} \\
b_2=\frac{1}{M^2}\left[a^2(E^2-1)-L_z^2-Q\right]=a_2, \label{10} \\
b_4=-\left(\frac{a}{M}\right)^2(E^2-1)=-\left(\frac{a}{M}\right)^2a_4,  \label{11}
\end{eqnarray}
where a dot stands for differentiation with respect to the dimensionless proper time $\tau$, 
and $E$, $L_z$ and $Q$ are integration constants of motion. Here Chandrasekhar's particle mass $\sqrt{\delta_1}$ \cite{Chandra}
has been set to one, its value for timelike geodesics, and
$E$ and $L_z$ have the usual significance of total energy and angular momentum
about the $z$ axis, both quantities given by units of $\sqrt{\delta_1}$, and $Q$ is the Carter constant given by units of $\delta_1$. With this
understanding, $E$ is dimensionless, while $L_z$ and $Q$ have dimensions of $M$ and $M^2$,
respectively. All the $a_i$ and
$b_i$ are dimensionless. In this paper we consider only particles moving along  unbound geodesics with $E\geq 1$ \cite{Chandra}.

In order to understand better the physical meaning of the Kerr geodesics we
use Weyl cylindrical coordinates $\rho $, $z$ and $\phi $ which are more
revealing and, in a way, a natural choice for axially symmetric
systems \cite{Gariel,Gariel1,Gariel2}. The dimensionless Weyl cylindrical coordinates, in multiples of
geometrical units of mass $M$, are given by
\begin{equation}
\rho =\left[(r-1)^2-A\right]^{1/2}\sin \theta ,\;\;z=(r-1)\cos\theta ,
\label{12}
\end{equation}
where
\begin{equation}
A=1-\left(\frac{a}{M}\right)^2.  \label{13}
\end{equation}
From (\ref{12}) we have the inverse transformation
\begin{eqnarray}
r=\alpha +1,  \label{14} \\
\sin\theta=\frac{\rho }{(\alpha ^2-A)^{1/2}},\;\;\cos\theta =\frac{z}{\alpha},  \label{15}
\end{eqnarray}
with
\begin{equation}
\alpha =\frac{1}{2}\left\{\left[\rho ^2+(z+\sqrt{A})^2\right]^{1/2}+
\left[\rho^2+(z-\sqrt{A})^2\right]^{1/2}\right\} .  \label{16}
\end{equation}
Here we have assumed $A\geq 0$, and taken the root of the second degree
equation obtained from (\ref{15}) for the function $\alpha (\rho ,z)$ that
allows the extreme black hole limit $A=0$. The other root in this limit is $\alpha =0$.

Now, with (\ref{14}) and (\ref{15}) we can rewrite the geodesics (\ref{2})
and (\ref{3}) in terms of $\rho $, $z$ and $t$ coordinates, producing the
following autonomous system of first order equations
\begin{eqnarray}
{\dot{\rho}}&=&\frac{1}{U}\left[\frac{P\alpha^3\rho}{\alpha ^2-A}+
 \frac{S(\alpha^2-A)z}{\alpha\rho}\right] ,  \label{17} \\
M{\dot{z}}&=&\frac{1}{U}\,(Pz-S)\alpha ,  \label{18}\\
{\dot{t}}&=&\frac{\alpha^2}{U\Delta }\left[\Sigma ^{2}E-2\frac{a}{M}(\alpha+1)\frac{L_z}{M}\right],  \label{Tdot}
\end{eqnarray}
where
\begin{widetext}
\begin{eqnarray}
P&=&\left[a_4(\alpha +1)^4+a_3(\alpha+1)^3+a_2(\alpha+1)^2+a_1(\alpha+1)+a_0\right]^{1/2},  \label{19} \\
S&=&-(b_4z^4+b_2\alpha^2z^2+b_0\alpha^4)^{1/2},  \label{20}\\
U&=&(\alpha+1)^2\alpha^2+\left(\frac{a}{M}\right)^2z^2,\label{21}\\
\Delta &=&(\alpha +1)^2-2(\alpha+1)+\left(\frac{a}{M}\right)^2=\alpha^2+\left(\frac{a}{M}\right)^2-1=\alpha^2-A,  \label{Delta}\\
\Sigma^2&=&\left[(\alpha+1)^2+\left(\frac{a}{M}\right)^2\right]^2+\left(\frac{a}{M}\right)^2
\left[\left(\frac{z}{\alpha}\right)^2-1\right]\Delta,  \label{Sigma2}
\end{eqnarray}
\end{widetext}
with the sign of $S$ chosen to indicate outgoing particles  \cite{Gariel}.

Our aim is to study geodesics that can attain large distances while
becoming parallels to the axis $z$, which corresponds to the condition
$z\gg \rho $. For this limit,   from (\ref{20}) we find
\begin{equation}
S\approx -\left[ (b_0+b_2+b_4)z^4+(2b_0+b_2)\rho ^2z^2\right]
^{1/2}+O(z^{-1}),  \label{30}
\end{equation}%
where
\begin{eqnarray}
b_0+b_2+b_4=-\left(\frac{L_z}{M}\right) ^2\leq 0,  \label{31}
\\
2b_0+b_2=\frac{1}{M^2}\left[ a^2(E^2-1)-L_z^2+Q\right] .
\label{32}
\end{eqnarray}
Hence in this limit $S$ is well defined and real for indefinitely small
$\rho /z$ only for $L_z=0$. The geodesics obeying this restriction, imposed
after similar reasoning, were studied in \cite{Bicak}, but in
Boyer-Lindquist or Kerr-Schild coordinates.

From now on we consider that $a$ and $L_z$ are also expressed in units of $M$ and $Q$ in units of $M^2$, which is equivalent to put $M=1$ in $a/M$,
$L_z/M$ and $Q/M^2$.

\section{Proper acceleration along the $z$ axis}

We would like now to calculate the proper acceleration $\ddot z$ of a particle along the $z$ axis.
Taking the second proper time derivative of (\ref{18}) produces (see \cite{Gariel2})
\begin{eqnarray}
U^2{\ddot{z}}&=&U({\dot{P}}z-{\dot{S}})\alpha +\left[ UP-2\left( \frac{a}{M}\right)^2(Pz-S)z\right] \alpha {\dot{z}}
\nonumber \\
&& +(Pz-S)\left[ U-2\alpha ^2(\alpha +1)(2\alpha +1)\right] {\dot{\alpha}}.\label{23}
\end{eqnarray}
From (\ref{7}), (\ref{19}) and (\ref{20}) we have
\begin{widetext}
\begin{eqnarray}
U({\dot{P}}z-{\dot{S}})=\frac{1}{2}\left[4a_4(\alpha+1)^3+3a_3(\alpha +1)^2+2a_2(\alpha+1)+a_1\right]\alpha^2z 
-2PS\alpha+(2b_4z^2+b_2\alpha^2)\alpha z.  \label{24}
\end{eqnarray}
Substituting (\ref{18})  and (\ref{24}) into (\ref{23}) we obtain
\begin{eqnarray}
\frac{U^3}{\alpha^2}\;{\ddot{z}}&=&
\frac{U}{2}\left[4a_4(\alpha +1)^3+3a_3(\alpha +1)^2+2a_2(\alpha +1)+a_1\right]\alpha z 
+U\left[(2b_4z^2+b_2\alpha^2)z-2PS\right]\nb\\
&& +2(Pz-S)\left[UP-\left(\frac{a}{M}\right)^2(Pz-S)z\right] 
-2(Pz-S)P\alpha^2(\alpha +1)(2\alpha +1),  \label{25}
\end{eqnarray}
or
\begin{eqnarray}
\frac{U^3}{\alpha^2}\;{\ddot{z}}&=&\frac{U}{2}\left[4a_4(\alpha+1)^3+3a_3(\alpha +1)^2+2a_2(\alpha+1)+a_1\right]
\alpha z 
+U(2b_4z^2+b_2\alpha^2)z-2P^2\alpha^3(\alpha +1)z\nb\\
&& -2\left(\frac{a}{M}\right)^2S^2z-2PS\alpha^2(\alpha +1).
\label{26}
\end{eqnarray}
Substitution of (\ref{19}), (\ref{20}) and (\ref{21}) into (\ref{25}) gives
\begin{eqnarray}
\frac{U^3}{\alpha^2}\;{\ddot{z}}&=& -\frac{1}{2}\left[a_3(\alpha+1)^3+2a_2(\alpha+1)^2+3a_1(\alpha+1)+4a_0\right]
\alpha^3(\alpha +1)z 
+\frac{1}{2}\left(\frac{a}{M}\right)^2\left[4a_4(\alpha+1)^3+3a_3(\alpha+1)^2\right.\nb\\
&& \left. +2a_2(\alpha+1)+a_1\right]\alpha z^3
+(2b_4z^2+b_2\alpha^2)\alpha^2(\alpha+1)^2z-\left(\frac{a}{M}\right)^2(b_2z^2+2b_0\alpha^2)\alpha^2z
-2PS\alpha^2(\alpha+1).  \label{27}
\end{eqnarray}
Now considering (\ref{4}), (\ref{7}), (\ref{10}) and (\ref{11}) and then 
substituting them into (\ref{27}) we have
\begin{eqnarray}
\frac{U^3}{\alpha^3}\;{\ddot{z}}&=& -(\alpha+1)^4\alpha^2z-
\left[\left(a_2+\frac{3}{2}\;a_1\right)(\alpha+1)^2+2a_0\right]\alpha^2z 
+\left(\frac{a}{M}\right)^2\left[(2a_4+3)(\alpha+1)^2+a_2+\frac{1}{2}a_1\right]z^3\nb\\
&& -2PS\alpha (\alpha+1).  \label{28}
\end{eqnarray}
\end{widetext}
Restoring the dimensions leads the proper $a_z$ component of the acceleration to the form
\begin{equation}
a_z=\frac{c^2}{M}{\ddot z}. \label{28b}
\end{equation}

\section{Proper acceleration along the $\rho$ axis}
In the same way as in the previous section, we compute the dimensionless proper acceleration $\ddot\rho$ along $\rho$ by
differentiating (\ref{17}), from which   we obtain
\begin{equation}
{\ddot\rho}=\frac{\alpha^2\Delta\rho^2R_2-2R_1}{2\alpha^2\Delta^2U^2\rho^3},
\label{29}
\end{equation}
with
\begin{widetext}
\begin{eqnarray}
R_1&=& [\alpha^4\rho^2P+z\Delta^2S] 
\left\{3\alpha^4\rho^2P+z\Delta S
+\frac{2\alpha^2\Delta\rho^2}{U}[\alpha^2(2\alpha^2+3\alpha+1)P+a^2z(Pz-S)]\right\},  \label{R1} \\
R_2&=& \alpha^5\rho^2\{a_1+r[2a_2r(3a_3+4ra_4)]\} 
+8\alpha^4\rho^2\{a_0+r[a_1+r(a_2+r(a_3+a_4r))]\} 
+8z\alpha^2\Delta PS+2\Delta^2S(Pz-S)\nb\\
&& +\frac{4\alpha^2P}{\Delta }(\alpha^4\rho^2P+Sz\Delta^2) 
+\frac{2z\Delta^2}{S}\{2\alpha^4Pb_0+z[2z^2b_4(Pz-S)+\alpha^2]b_2(2Pz-S)\}.  \label{R2}
\end{eqnarray}
\end{widetext}
Restoring the dimensions leads  the proper $a_{\rho}$ component of the acceleration to
\begin{equation}
a_{\rho}=\frac{c^2}{M}{\ddot\rho}. \label{R2A}
\end{equation}

\section{$t$ acceleration}

In the same way, we compute the t acceleration, i.e. the dot derivative of
$\dot t$ given by (\ref{Tdot}). In fact, we need
$(1/\dot t){\dot{}}$, as we shall see in section VI. We obtain, by limiting
ourselves to the case $L_{z}=0$, which is interesting to us, 
\begin{equation}
\left(\frac{1}{\dot t}\right)^{\dot{}}=\frac{2\alpha ^{3}T_{1}}{EUT_{2}}, \label{Dot1surTdot}
\end{equation}
with
\begin{widetext}
\begin{eqnarray}
T_1&=& P\alpha^3r^2[(a^2-1)^2+4\alpha+2(a^2+3)\alpha^2+4\alpha^3+\alpha^4] 
-2a^2Szr[-1+a^4+2(a^2-1)\alpha+2a^2\alpha^2+2\alpha^3\alpha^4] \nonumber   \\
&& -a^2Pz^2\alpha\lbrack-3+2a^2a^4+4(a^2-2)\alpha+2(a^2-3)\alpha^2+\alpha^4],  \label{T1} \\
T_2&=& \{a^2z^2\Delta+\alpha^2r[1+3a^2+(3+a^2)\alpha+3\alpha^2+\alpha^3]\}^2.  \label{T2}
\end{eqnarray}
\end{widetext}

\section{Observable velocity and acceleration}

We define ``observable velocity and acceleration," the set of components of the velocity and acceleration (in our case along $z$ and $\rho$),
 as measured with the time $t$. These velocities and accelerations are observable at infinity, where spacetime is Minkowskian.

The time $\hat t$, the coordinate $\bar t$, which has dimensions of length like $\bar\rho$ and $\bar z$ (see \cite{Chandra}), and the dimensionless coordinate $t$ are linked by the relations
\begin{equation}
{\hat t}=\frac{\bar t}{c}=\frac{M}{c}t. \label{T2a}
\end{equation}

The $v_z$ velocity is given by
\begin{equation}
v_z=\frac{d{\bar z}}{d{\hat t}}=\frac{Mdz}{(M/c)dt}=c\frac{dz}{dt}=c\frac{\dot z}{\dot t}, \label{vz}
\end{equation}
where $\dot z$ and $\dot t$ are given by (\ref{18}) and (\ref{Tdot}), respectively. We observe that $v_z$ is a function of $\rho$ and $z$ and in the limit $z\rightarrow\infty$ we have
\begin{equation}
\beta_z=\frac{v_z}{c}\rightarrow\left(1-\frac{1}{E^2}\right)^{1/2}, \label{vz1}
\end{equation}
which is the Minkowskian expression mentioned in \cite{Gariel} equation (49) where we recognize that $E$ is the Lorentz factor $\Gamma=(1-\beta^2)^{-1/2}$ when $\beta_z\rightarrow\beta$, i.e., $\beta_{\rho}\rightarrow 0$.
The dimensionless acceleration along $z$ is
\begin{equation}
\beta_z^{\prime}=\frac{d\beta_z}{dt}=\frac{1}{\dot t}\left(\frac{\dot z}{\dot t}\right)^{\dot{}}. \label{az}
\end{equation}
Restoring the dimensions, we find that  the   component of acceleration along $z$-direction is given by
\begin{equation}
\gamma_z=\frac{dv_z}{d\hat t}=c\frac{d\beta_z}{d\hat t}=\frac{cd\beta_z}{(M/c)dt}=\frac{c^2}{M}\beta_z^{\prime}. \label{Gamz}
\end{equation}

Similarly, the formulae for the velocity and acceleration along $\rho$-direction  are, 
\begin{eqnarray}
v_{\rho}=c\frac{\dot\rho}{\dot t}=c\beta_{\rho}, \label{Vrho}\\
\beta_{\rho}^{\prime}=\frac{d\beta_{\rho}}{dt}=\frac{1}{\dot t}\left(\frac{\dot\rho}{\dot t}\right)^{\dot{}}, \label{Arho}
\end{eqnarray}
where
\begin{equation}
\gamma_{\rho}=\frac{dv_{\rho}}{d\hat t}=\frac{c^2}{M}\beta_{\rho}^{\prime}. \label{Gamrho}
\end{equation}

The conversion coefficient $c^{2}/M$ from the dimensionless to the
dimensional observable accelerations (\ref{Gamz}) and (\ref{Gamrho}) is the same as for the proper
accelerations in (\ref{28b}) and (\ref{R2A}), and  for the M87 galaxy black hole we find 
\begin{equation}
\frac{c^{2}}{M}=\frac{9\time 10^{16}}{6\times 10^{9}\times 1.5\times 10^{3}}=10^{4}\, \mbox{ms}^{-2}, \label{X1}
\end{equation}
for $M=6\times 10^{9}M_{\odot}$, in accordance with recent evaluations \cite{Gebhardt}.

Now we have all the quantities  for the observable components of the velocity
and acceleration as functions of $\rho $ and $z$. Along the geodesics these
two variables are no long independent, instead, $\rho $ now is a function of $z$, which
describes the trajectory that a particle follows for the given
geodesic. The $z$ and $\rho $ components of the proper and observable
velocities and accelerations now can be evaluated along the
geodesic as functions of $z$ only.

\section{Acceleration for high energy jets}

The trajectories followed by the particles running along the geodesics are
obtained by eliminating the time between the equations of motion (\ref{17})
and (\ref{18}), i.e.,  they are the solutions of the differential equation,
\begin{equation}
\frac{d\rho}{dz}=\frac{P\alpha^4\rho^2+S(\alpha^2-A)^2z} {%
(Pz-S)(\alpha^2-A)\alpha^2\rho}.  \label{Traject}
\end{equation}

We consider special geodesics with trajectories given by (\ref{Traject}). When $z\rightarrow\infty$,
we find that  $\rho \rightarrow\rho_1$, with the asymptotes
\begin{equation}
\rho_1=\rho_e\left[1+\frac{Q}{a^2(E^2-1)}\right]^{1/2},  \label{Rho1}
\end{equation}
where $\rho_e=a/M$, highlighted in \cite{Gariel} equation (21), to account
for a perfect collimation of jets. Such geodesics need that $L_z=0$, as seen
after equation (32) in \cite{Gariel}. This new parameter, $\rho_1$, allows a
new indirect interpretation of the Carter constant $Q$. Indeed, $\rho_1$ is
a parameter characterizing  the collimated jet's ejection.

We need to find the components (\ref{28}) and (\ref{29}), in order to obtain
the observable components, (\ref{Gamz}) and (\ref{Gamrho}), of the
acceleration as functions of $z$. In principle this is possible since
from (\ref{Traject}) one can obtain $\rho $ as a function of $z$, given initial
conditions $\{\rho _{i},z_{i}\}$. The same can be achieved for the
observable velocities, (\ref{vz}) and (\ref{Vrho}), given as functions 
of $z$ only. However such solutions can be determined only numerically. We can
fix the parameters $\rho _{1}$ and $E$, specially for high $E$, which
corresponds to the conspicuous part of the jet, as seen in \cite{Barba}.

As an example, let us take the following geodesic determined by
$\rho_{1}=3.16354$, $E=10^{6}$, $a/M=0.877004$, and by the initial conditions
inside the ergosphere, $\rho _{i}=0.5648$, $z_{i}=-0.25$, which is the
geodesic studied in section 4 of \cite{Barba}, i.e.
asymptotically parallel to the $z$ axis along $\rho =\rho _{1}$ with high
energy,  which is central for the M87 jet in our model (See figure 1). The trajectory of this geodesic is numerically
determined by numerical integration of the differential equation (\ref{Traject}), giving us an interpolating function $\rho =f(z)$ that we can
substitute into the different precedent expressions of the velocity and
acceleration, which leads us to the following results.

\begin{figure}[tbp]
\centering
\includegraphics[width=8.5cm]{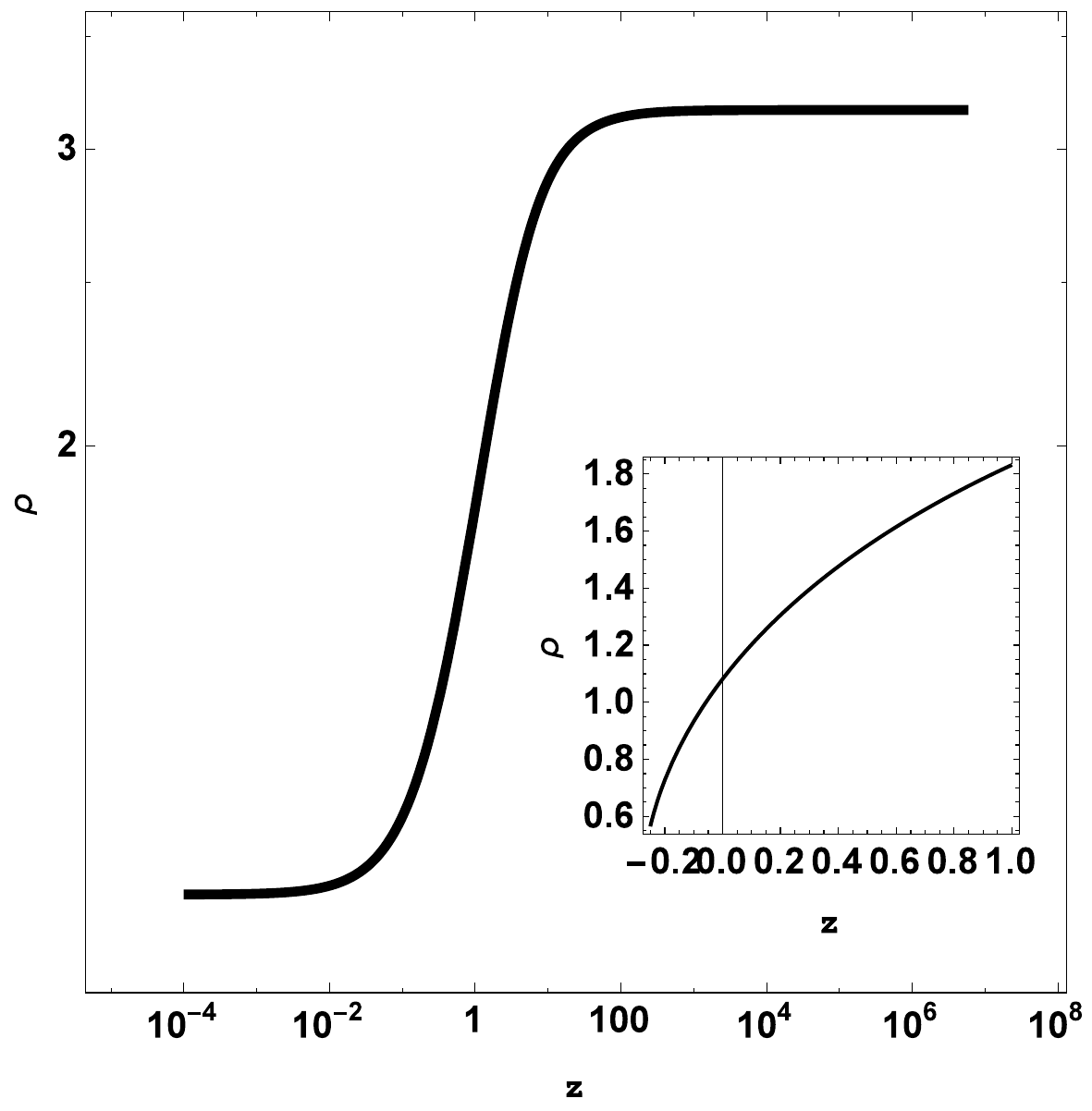}
\caption{LogLogPlot of the geodesic obtained by numerical integration
of (51) for the parameters $a=0.877$, $E=10^{6}$, asymptotic to $\rho
_{1}=3.16354$ with the initial condition $\rho _{i}(z_{i}=-0.25)=0.5648$,
 for the range $z\in \lbrack 10^{-4},6\times 10^{6}]$ , and
(inserted inside) linear plot of the beginning for $z\in \lbrack -0.25,1]$.
}
\label{Fig1}
\end{figure}

\subsection{Velocity}

The interpolating function $\rho (z)$ used in (\ref{vz}) and (\ref{Vrho}),
yields a numerical solution for the observable velocity components $\beta_{z}(z)$ and $\beta _{\rho }(z)$ and norm $\beta (z)$ along this geodesic,
which are plotted in figure 2. Table 1 summarizes
the main numerical results. Comments on these curves and table 1 are the
following.

 \begin{widetext}
$$
\begin{tabular}{llllll}
$z$ & $\overline{z}$ (pc) & $\beta _{z}$ & $\beta _{\rho }$ & $\beta $ & $%
\gamma $ \\
$-0.25$ & $-0.75\times 10^{-4}$ & $0.0264676$ & $0.113153$ & $0.116207$ & $%
\allowbreak 1.\,\allowbreak 006\,8$ \\
$0.033$ & $10^{-5}$ & $0.153425$ & $0.19038$ & $0.244441$ & $\allowbreak
1.\,\allowbreak 031\,3$ \\
$0.5$ & $1.5\times 10^{-4}$ & $0.284537$ & $0.196846$ & $0.345991$ & $%
\allowbreak 1.\,\allowbreak 065\,8$ \\
$0.75$ & $\allowbreak 2.\,\allowbreak 25\times 10^{-4}$ & $0.336773$ & $%
0.188696$ & $0.386034$ & $\allowbreak 1.\,\allowbreak 084$ \\
$2$ & $6\times 10^{-4}$ & $0.510814$ & $0.134569$ & $0.528242$ & $%
\allowbreak 1.\,\allowbreak 177\,7$ \\
$4.25$ & $\allowbreak 1$.$2\,\allowbreak 4\times 10^{-3}$ & $0.670313$ & $%
0.0713721$ & $0.674102$ & $\allowbreak 1.\,\allowbreak 353\,8$ \\
$5$ & $1.5\times 10^{-3}$ & $0.7036$ & $0.0591092$ & $0.706079$ & $%
\allowbreak 1.\,\allowbreak 412\,2$ \\
$10$ & $3\times 10^{-3}$ & $0.825586$ & $0.0220434$ & $0.825881$ & $%
\allowbreak 1.\,\allowbreak 773\,5$ \\
$20$ & $6\times 10^{-3}$ & $0.90593$ & $0.00668179$ & $0.905955$ & $%
\allowbreak 2.\,\allowbreak 362\,0$ \\
$33$ & $10^{-2}$ & $0.942018$ & $0.00258397$ & $0.942021$ & $\allowbreak
2.\,\allowbreak 980\,1$ \\
$50$ & $1.5\times 10^{-2}$ & $0.96087$ & $0.00118779$ & $0.96087$ & $%
3.\,\allowbreak 610\,1$ \\
$70$ & $2.1\times 10^{-2}$ & $0.971863$ & $6.175\times 10^{-4}$ & $0.971863$
& $\allowbreak 4.\,\allowbreak 245\,4$ \\
$100$ & $0.03$ & $0.980209$ & $3.07\times 10^{-4}$ & $0.980209$ & $%
\allowbreak 5.\,\allowbreak 051\,4$ \\
$333$ & $0.1$ & $0.994018$ & $2.8\times 10^{-5}$ & $0.994018$ & $%
9.\,\allowbreak 156\,1$ \\
$500$ & $0.15$ & $0.996008$ & $1.2\times 10^{-5}$ & $0.996008$ & $%
\allowbreak 11.\,\allowbreak 203$ \\
$10^{3}$ & $0.3$ & $0.998002$ & $3.15\times 10^{-6}$ & $0.998002$ & $%
\allowbreak 15.\,\allowbreak 827$ \\
$10^{4}$ & $3$ & $0.9998$ & $3.16\times 10^{-8}$ & $0.9998$ & $\allowbreak
50.\,\allowbreak 003$ \\
$10^{5}$ & $30$ & $0.99998$ & $3.1\times 10^{-10}$ & $0.99998$ & $%
\allowbreak 158.\,\allowbreak 11$ \\
$10^{6}$ & $300$ & $0.9999980$ & $10^{-11}$ & $0.999998$ & $\allowbreak 500.0
$%
\end{tabular}
$$
\textbf{Table 1}. \ Some indicative values of the observable velocity (components $\beta _{z}$, $\beta _{\rho }$ and norm $\beta $) and Lorentz
factor ($\gamma $) as functions of $z$ along the geodesic asymptotic to $\rho _{1}=3.1635$ for a particle of observable energy $E=10^{6}$ (in units of its mass)
with initial conditions $\rho _{i}(z_{i}=-0.25)=0.5648$ inside the Kerr ergosphere of the M87 BH of mass $M=6\times 10^{9}M_{S}$. See figure 2.
The initial values are $\beta _{zi}=0.026$ and $\beta _{\rho i}=0.113153$.
These initial observable velocities are non relativistic (NR).

\end{widetext}

\begin{figure}[tbp]
\includegraphics[width=1\columnwidth,height=7.15cm]
{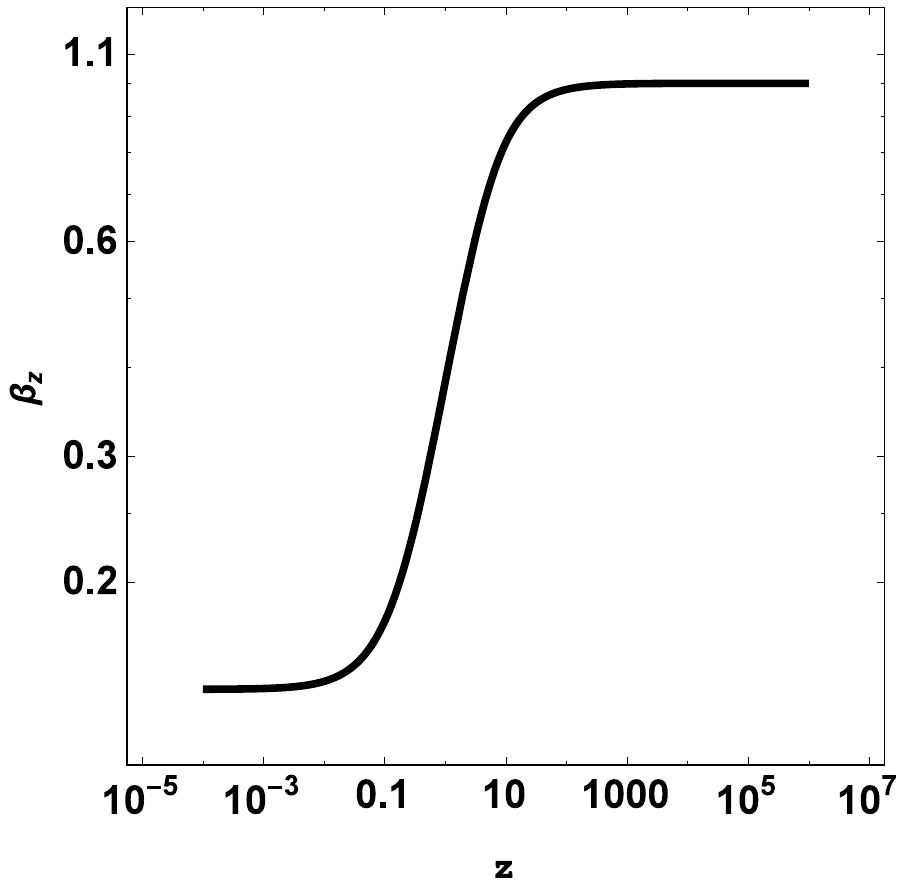}\\
\includegraphics[width=1\columnwidth,,height=7.15cm]{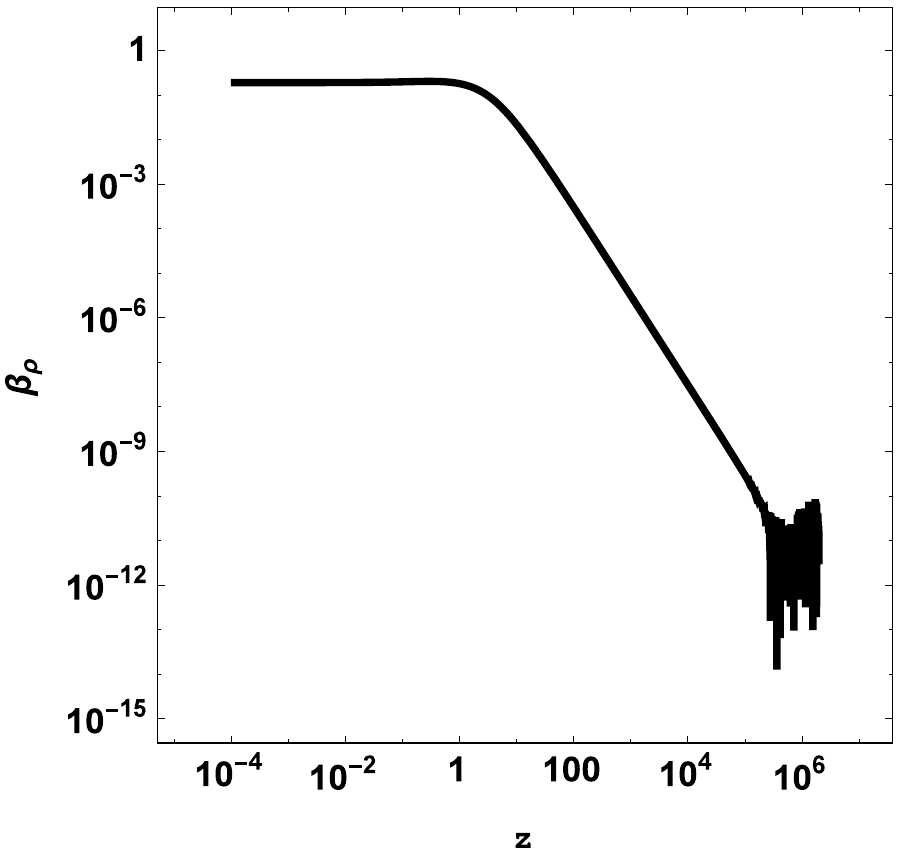}\\
\includegraphics[width=1\columnwidth,,height=7.15cm]{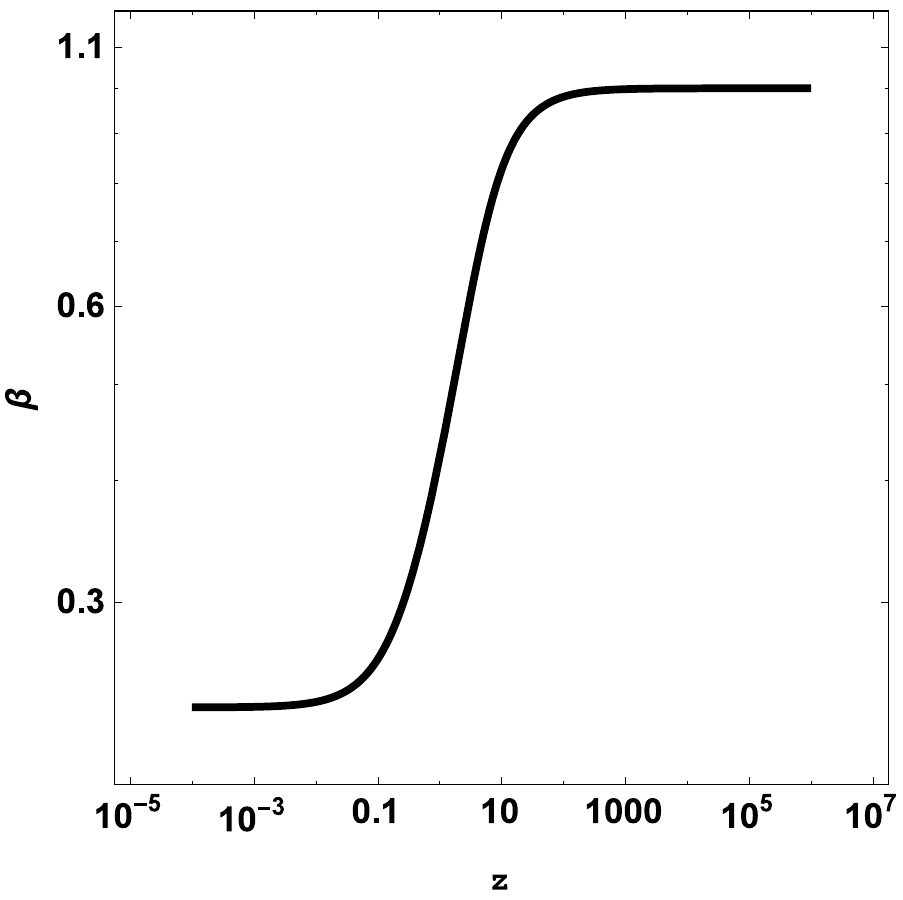}
\caption{Plots in LogLog scale of the observable velocity in
function of $z$ along the geodesic plotted in figure 1. Components $\beta
_{z}=v_{z}/c$ \ (43), $\beta _{\rho }=v_{z}/c$ (47) and norm $\beta =(\beta
_{z}^{2}+\beta _{\rho }^{2})^{1/2}$.
}
\label{Fig2}
\end{figure}

When $z=10^{6}$ for M87, taking the recent value obtained for $M$
\cite{Gebhardt}, that corresponds to $\overline{z}=10^{6}M\simeq 10^{6}\times 6\times 10^{9}\times M_{\odot}=6\times 10^{15}\times \frac{3}{2}$ km $=9\times
10^{15}$ km $=9\times 3.241\times 10^{15}\times 10^{-14}$ pc $=291.69$ pc $\simeq 300$ pc, the values are $\beta _{z}=0.999998$ and $\beta _{\rho }\approx 10^{-11}$. 
At the infinity, the observable Lorentz factor tends to its  maximum possible value $\gamma _{\infty }=\sqrt{E^{2}-1}\simeq E=10^{6}$ (see equations (47) and (48) in \cite{Gariel}, and figure 5 in \cite{Pacheco}),
which is ultra-relativistic (UR). For an electron or a proton, it
corresponds to the energy $\sqrt{\delta _{1}}\gamma \simeq 0.5\times
10^{12}$ eV, or $10^{15}$ eV, respectively.

We observe that $\beta _{\rho }$\ has a (very) weakly relativistic maximum
$\approx 0.2$ for $z\simeq 0.5$, then rapidly becomes non
relativistic and tends to zero, while the component $\beta _{z}$\ becomes
relativistic, $\beta _{z}\sim 0.8$, near $z\sim 10$, and ultra relativistic,
$\beta _{z}\sim 0.98$,\ near $z\sim 100$. The curve figure 2c shows that $\beta (z)$
becomes rapidly very close to the curve presented in  figure 2a for $\beta _{z}(z)$.

These results (specially the shape of the curve $\beta _{z}(z)$ and the
limit of $\beta _{z}$ when $z\rightarrow \infty $) qualitatively remain in
accordance with those obtained in \cite{Pacheco} from different conditions
($a=0.5$,  $Q<0$).

We remark that the shape of the curve in figure 2a (or 2c) is very similar to those
obtained from recent observations (see the  bottom  part of figure 3 in \cite{Lee}),  though the
assumptions (involving a magnetic field) from which they interpret the
observations are very different from ours. However, their transition (from $\beta \sim 0.5$ NR to $\beta \sim 0.9$ UR) is located to about $r\in
\lbrack 0.3,1]$ pc, while ours, located about $z\in \lbrack 2,20]$,
i.e., $\overline{z}\in \lbrack 0.6,6]\times 10^{-3}$ pc, is more abrupt.
Besides, our model predicts that the phenomenon happens earlier (closer
to the core) than that happening in their model. In fact, from   figure 3 of  \cite{Lee}, one can see that the range is from $10^{-2}$ to
$10^{2}$ pc, with a transition between $10^{-1}$ and $10$ pc,  while in
our model, as can be seen  from our figure 2c,  the range runs  from about $10^{-4}$\ to $1$ pc, with a
transition between $10^{-4}$\ and $10^{-2}$ pc.  We shall come back to this issue in Section VIII.

Observations are done by the radiation emitted by the jet. It is accepted
that the main source of this radiation is synchrotron radiation, which is
continuous. What is observed is an almost continuous sequence of frequencies
increasingly elevated to as we ascend from the stream to its source, e.g.
\cite{Asada,Hada,Hada1}

Most of the methods for determining the acceleration of the jet are
model-dependent: they assume that the electromagnetic field plays the dual roles:  the
cause of the acceleration and the creation of  the synchrotron effect. But it
can be also considered a situation where the cause of the  accelerator  is
different from the origin of the synchrotron effect. This is the case of our
model, where the origin  of the acceleration is gravitational, while the
particle accelerated to some speed will enter a magnetic field, whereby 
generating the synchrotron radiation. The magnetic field acts like an
accelerator, but weak, and the energy gained by the particles  will be quickly  lost through  radiation, making inefficient the magneto-accelerator phenomenon
over long distances. The observed critical frequency is proportional to $B\gamma^2$. At the same $z$,  the
critical synchrotron frequency has a bigger Lorentz factor than that in the
magnetic models, which means that the magnetic field deduced form our model  is smaller
than the one deduced from the previous ones.    

\subsection{Acceleration}

The proper and observable accelerations are shown in figures \ref{Fig3a},  \ref{Fig3b},  \ref{Fig4a} and \ref{Fig4b},
respectively. Table 2 summarizes the main (numerical results). Some comments on
these curves and table 2 are in order.

\subsubsection{Proper acceleration}

\begin{figure}[tbp]
\includegraphics[width=8.5cm]{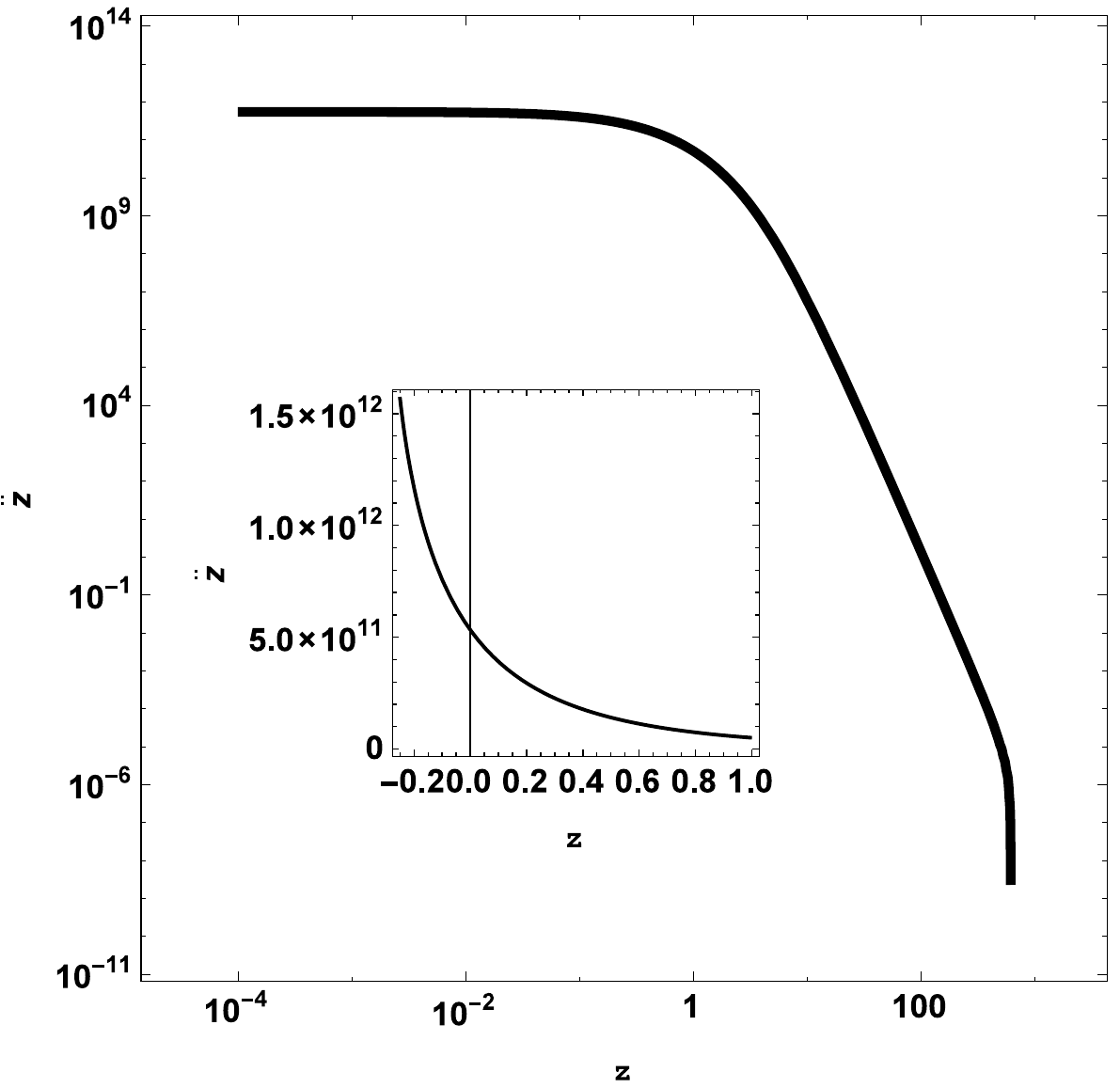}\\
\includegraphics[width=8.5cm]{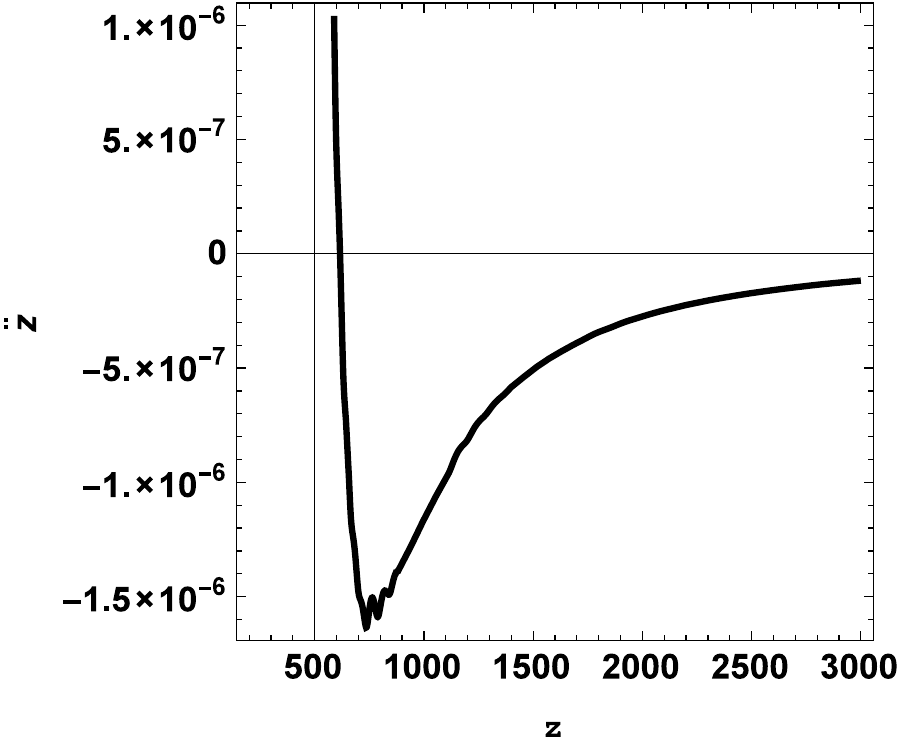}
\caption{ Plots of $\overset{\cdot \cdot }{z}$, the $z$-component
of the proper acceleration (33), along the geodesic of the figure1, in
function of $z$. General plot in LogLog scale and two local linear plots for
two different ranges : $z\in \lbrack -0.25,1]$, starting from a maximum
(inserted inside),\ and $z\in \lbrack 10^{2},3\times 10^{3}]$, showing the
minimum. There is a final increasing towards zero by negative values.
%
}
\label{Fig3a}
\end{figure}

\begin{figure}[tbp]
\includegraphics[width=1\columnwidth]{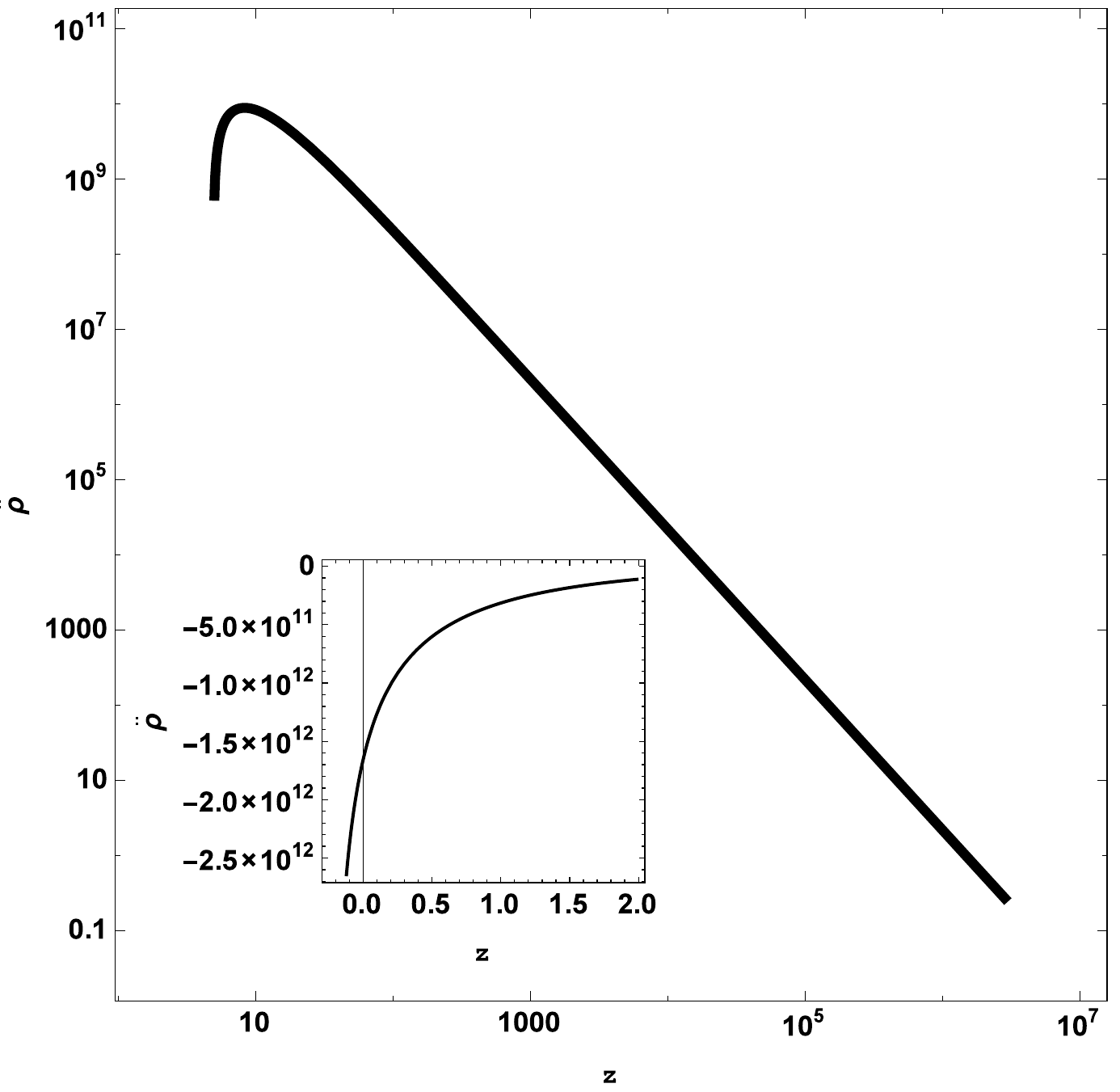}\\
\includegraphics[width=1\columnwidth]{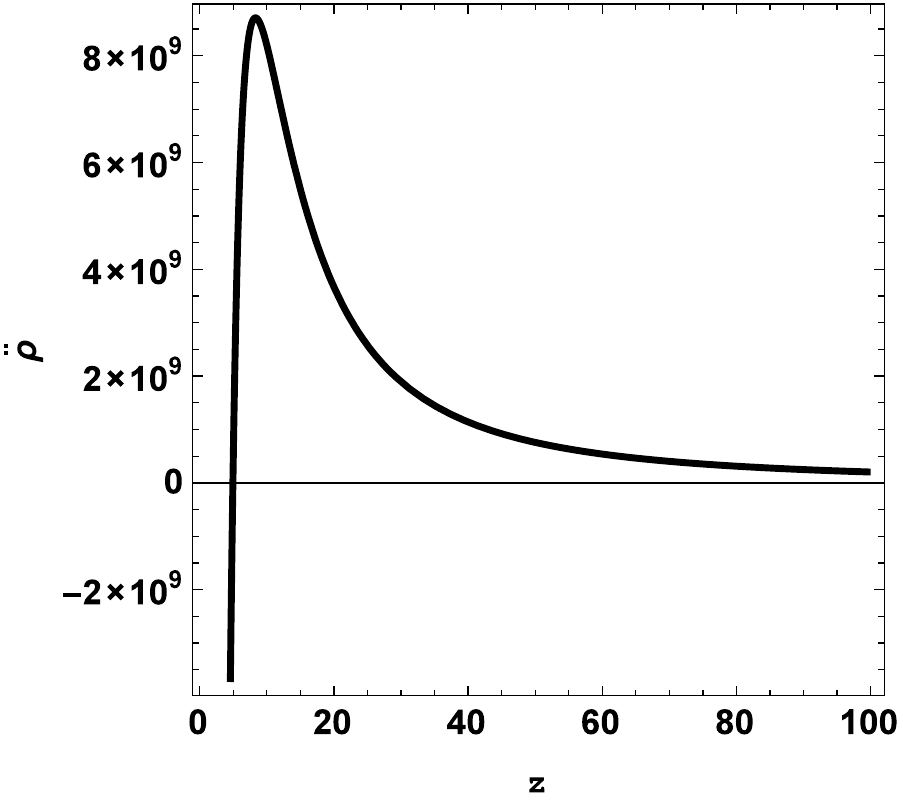}
\caption{Plots of $\overset{\cdot \cdot }{\rho }$, the $\rho $%
-component of the proper acceleration  (35), along the geodesic of the
figure 1, in function of $z$. General plot in LogLog scale and two local
linear plots for the following ranges of $z$ : $[-0.25,2]$ showing the
beginning (inserted inside), and $[1,10^{2}]$ showing the maximum. There is
a final decreasing towards zero by positive values.
%
}
\label{Fig3b}
\end{figure}

Introducing the interpolating function in the proper acceleration components
extracted from  equations (\ref{28}) and (\ref{29}), yields $\ddot{z}$
and $\ddot{\rho}$ as functions  of $z$ only along the geodesic, functions
that are  plotted in figures \ref{Fig3a} and \ref{Fig3b}, respectively.

{\bf (i)} Examining figure \ref{Fig3a}  in detail,  we have the following
remarks. The dimensionless proper acceleration $\ddot{z}$ starts from a large
positive initial value $\ddot{z}_{i}=\ddot{z}_{\max}=1.5755\times 10^{12}$  ($a_{zi}=1.57\times 10^{16}$ ms$^{-2}$), and then decreases
while remaining positive (repulsive force) until $z\simeq 625$ ($\overline{z}=625\times M=625\times 6\times 10^{9}\times 1.5$ km $=5.\,
\allowbreak 625\times 10^{12}$ km $=\allowbreak 0.182\,31$ pc), where it vanishes. Beyond that point, it becomes
negative (attractive force) until it reaches its smallest value $\ddot{z}
\simeq -1.4\times 10^{-6}$ ($a_{z}=-1.4\times 10^{-2} $ms$^{-2}$)\ at about $z\simeq 740$ ($\overline{z}=\frac{740}{625}\times 0.18231$ pc $=0.215\,86$ pc), 
before beginning to increase and tending to zero, while always remaining
negative. $\ddot{z}$\ reaches about \ $-10^{-7}$\ ($a_{z}=-10^{-3}$ ms$^{-2}$)
for $z\simeq 3\times 10^{3}$ ($\overline{z}=\frac{3000}{625}\times
0.18231$ pc $=\allowbreak 0.875\,09$ pc), and then \ $-2\times 10^{-9}$ ($a_{z}=-2\times 10^{-5}$ ms$^{-2}$), for $z\simeq 2.2\times 10^{4}$($\overline{z}=
\frac{2.2\times 10^{4}}{625}\times 0.18231$ pc $=\allowbreak 6.\,\allowbreak
417\,3$ pc). So, we can obtain the value of the (dimensionless) acceleration
component $\ddot{z}$ for each value of $z$ all along the geodesic. In
particular, $\ddot{z}\rightarrow 0$ when $z\rightarrow \infty $.

Since the initial position is the closest one to the BH, it is normal that the
initial proper acceleration is maximal, meaning a maximal repulsive force,
felt by the particle, in its proper (comoving) frame. The proper
acceleration in a sense is the expression of the strength of the
gravitational field. In Table 2  we present  some particular values of these physical quantities.

{\bf (ii)}  On the other hand, some remarks about  figure \ref{Fig3b} are the following. Initially, we have  $\ddot{\rho}_{i}=-8.60382\times 10^{12}$ ($a_{\rho i}=-8.6\times 10^{16}$ ms$^{-2}$). Then, it
increases to $\ddot{\rho}=-7.79134\times 10^{9}$\ ($a_{\rho }=-7.79\times 10^{13}$ ms$^{-2}$) at $z=4.25$\ ($\overline{z}=\frac{4.25}{625}\times 0.18231=\allowbreak 1.\,\allowbreak 239\,7\times 10^{-3}$ pc),
 and abruptly becomes positive $\ddot{\rho}=+9.89455\times 10^{7}$ ($a_{\rho }=9.9\times 10^{11}$ ms$^{-2}$) at $z=4.95$\ ($\overline{z}=\frac{4.95}{625}\times
0.18231=1.444\times 10^{-3}$ pc, where $a_{z}$\ is slightly greater than $a_{\rho }$).\ It vanishes between these two last positions, for $z\simeq
4.938173812$, from which\ it positively increases towards a maximum \ $\ddot{\rho}_{\max }=+8.71056\times 10^{9}$ ($a_{\rho }=8.7\times 10^{13}$ ms$^{-2}$)
at $z=8.308$ ($\overline{z}=\frac{8.308}{625}\times 0.18231$ pc $=2.\,\allowbreak 423\,4\times 10^{-3}$ pc). Then, it decreases towards zero
while remaining always positive.\ For the sake of comparison with $\ddot{z}$ we note that: at $z=30$, $\ddot{\rho }
=1.89417\times 10^{9}$\ (about $10^{6}$\ times greater than $\ddot{z}$)\ ; At $z=100$, $\ddot{\rho }=2.03259\times
10^{8}$\ (about $10^{8}$\ greater than $\ddot{z}$).
At $z=625$\ (where $\ddot{z}$\ vanishes), $\ddot{\rho }=5.48651\times 10^{6}$. At $z=740$\ (where $\ddot{z}$\ is minimum and negative $\sim -10^{-6}$),
$\ddot{\rho }=3.91956\times 10^{6}$; at $z=4\times 10^{3}$, $\ddot{\rho }=1.35023\times 10^{5}$. At $z=10^{4}$, $\ddot{\rho }
=2.16227\times 10^{4}$; at $z=3\times 10^{5}$,$\ \ddot{\rho }=20$; and at $z=10^{6}$, $\ddot{\rho }=2.16$. Finally,   we have $\ddot{\rho}\rightarrow 0$, 
 when $z\rightarrow \infty$.

\subsubsection{Observed acceleration}

\begin{figure}[tbp]
\includegraphics[width=1\columnwidth]{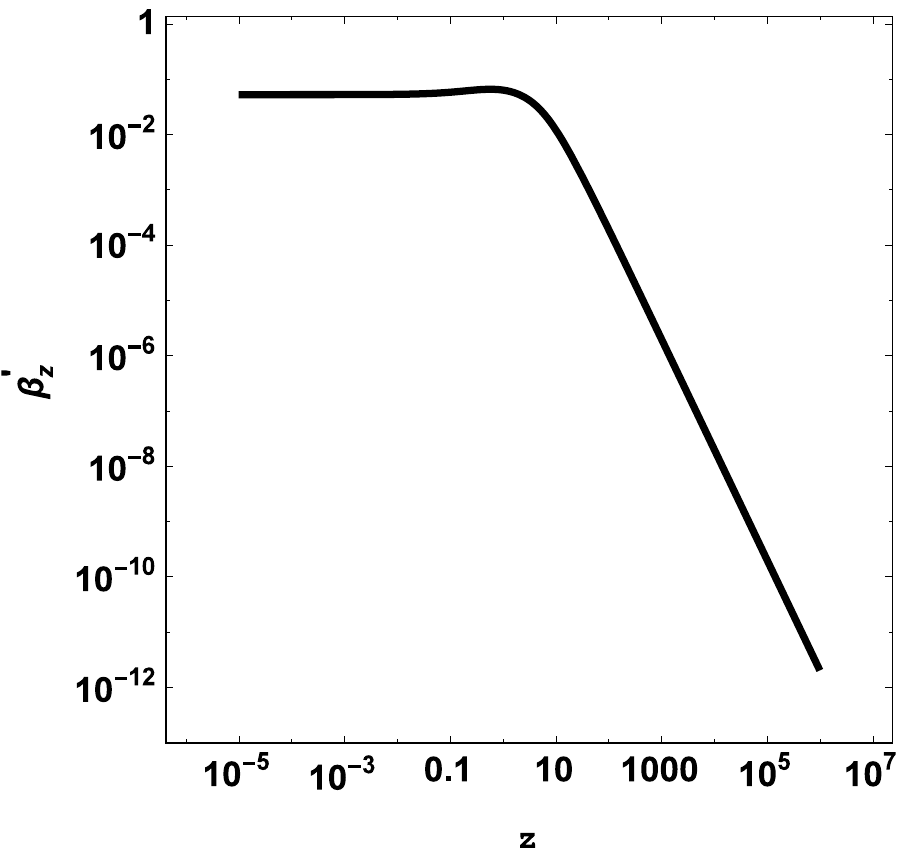}\\
\includegraphics[width=1\columnwidth]{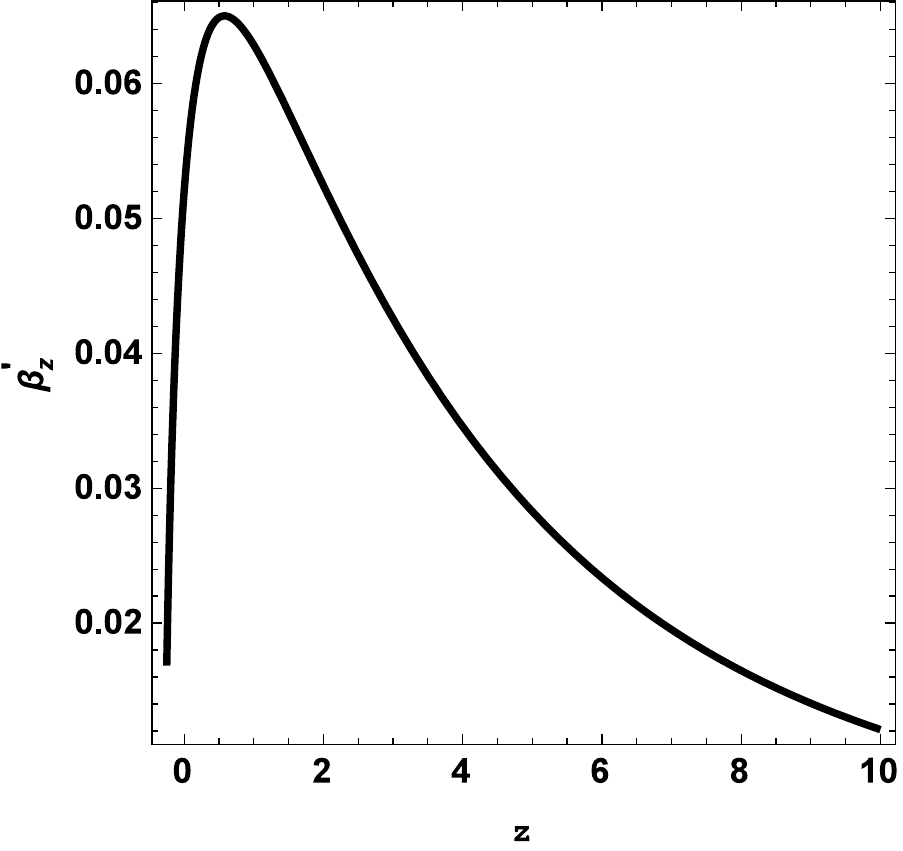}
\caption{
Plots of the $z$-component $\beta _{z}^{\prime }$ of the
observable acceleration  (45), along the geodesic of the figure 1, in
function of $z$. General plot in LogLog scale, and one local linear plot for
the $z$-range $[-0.25,10]$. Beyond, it decreases to zero. It is always
positive.
%
}
\label{Fig4a}
\end{figure}

\begin{figure}[tbp]
\includegraphics[width=1\columnwidth,height=7.15cm]{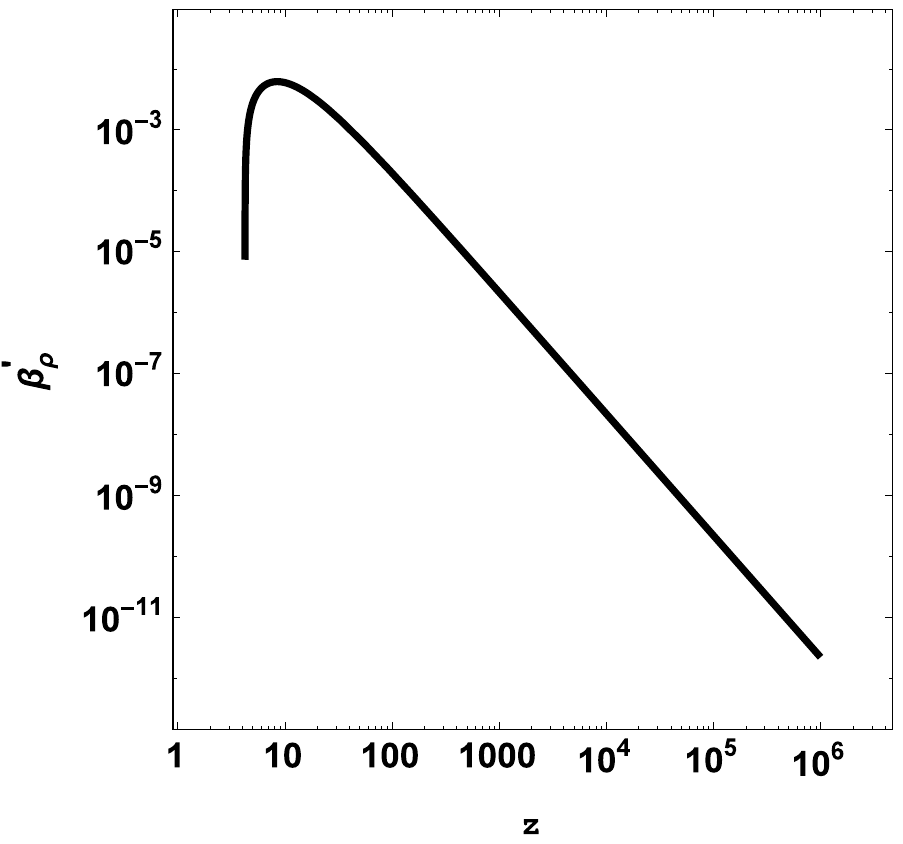}\\
\includegraphics[width=1\columnwidth,height=7.15cm]{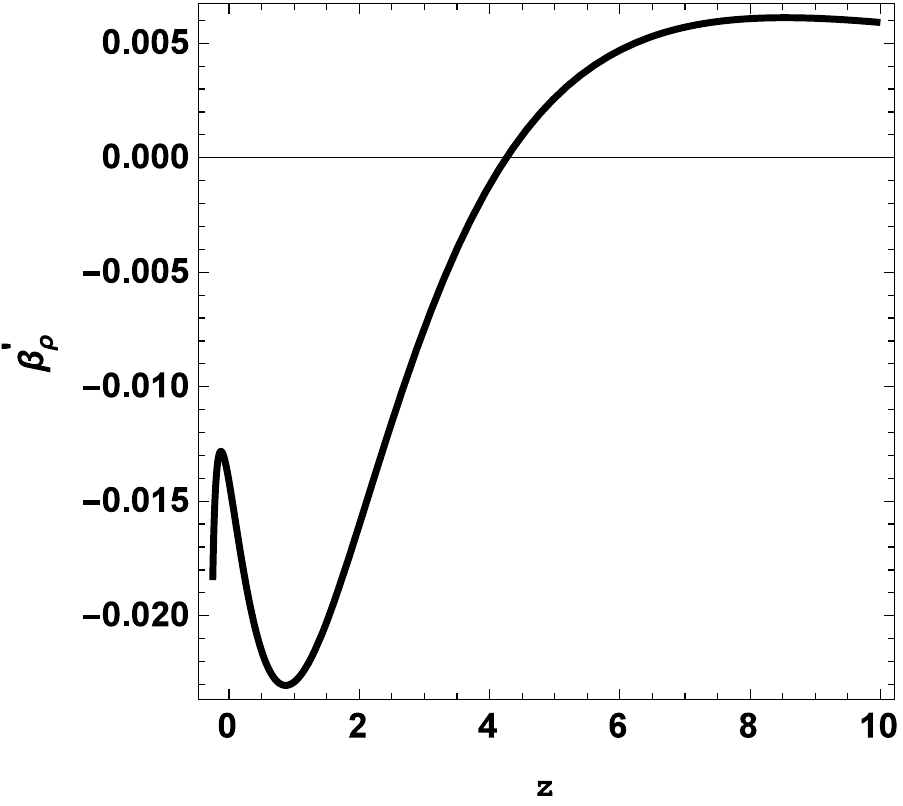}\\
\includegraphics[width=1\columnwidth,height=7.15cm]{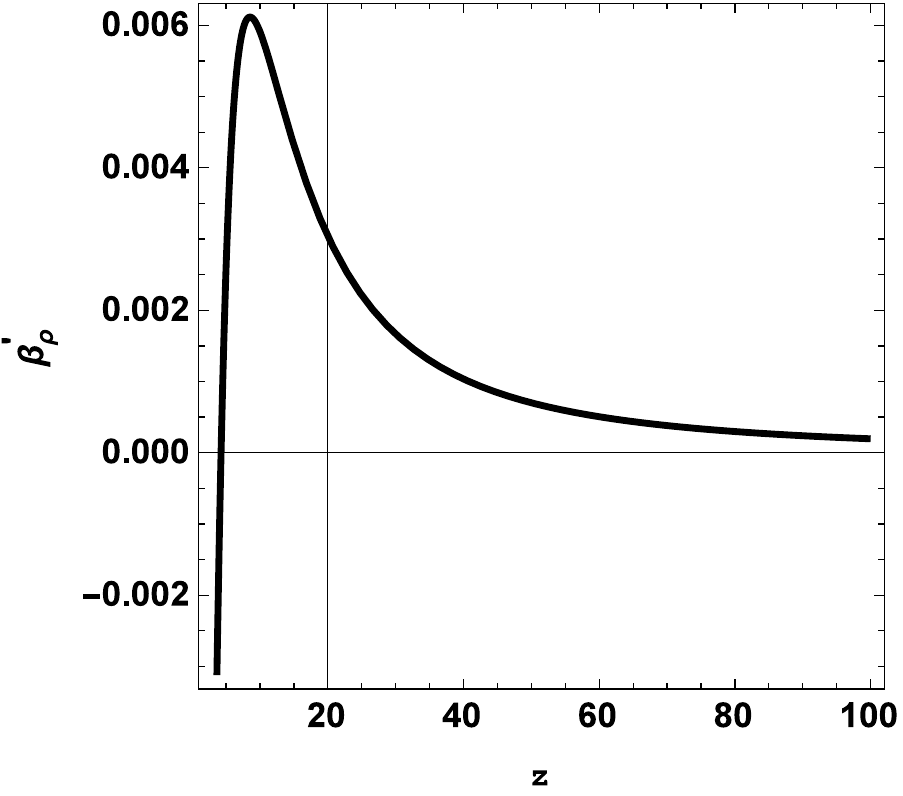}\\
\caption{Plots of the $\rho $-component $\beta _{\rho }^{\prime }$
of the observable acceleration (48), along the geodesic of the figure 1, in
function of $z$. General plot in LogLog scale, and two local linear plots
for the $z$-ranges $[-0.25,10]$ , showing the oscillation at the beginning,
and $[3,10^{2}]$ showing the maximum. There is a final decreasing to zero.
It is always positive beyond $z\sim 4.5$.
%
}
\label{Fig4b}
\end{figure}

We also plot the observed dimensionless acceleration components
$\beta_{z}^{\prime }$ of (\ref{az})  in figure \ref{Fig4a} and $\beta _{\rho }^{\prime }$ of (\ref{Arho})  in figure \ref{Fig4b}, by using the time of
the observer $t$,  instead of the proper time $\tau$. Then, from
(\ref{Gamz}) and (\ref{Gamrho}) we deduce the observed dimensional accelerations $\gamma_{z}$ and $\gamma _{\rho }$. In the following, let us consider these figures
in some detail. 

{\bf (i)} For $\beta _{z}^{\prime}$. We first note that it
is always positive. Its initial value, at $z_{i}=-0.25$, is \ $\beta_{zi}^{\prime }=0.0168428$, i.e., $\gamma _{zi}=168.428$ ms$^{-2}$. It
increases until its maximal value, $\beta _{z\max }^{\prime }=0.065$\ (or $\gamma _{z\max }=650$ ms$^{-2}$), obtained at $z=0.75$. Then, it decreases,
by remaining always positive. At $z=4.25$, $\beta _{z}^{\prime }=0.0328446$\
(or $\gamma _{z}=328.4$ ms$^{-2}$), at $z=10$, $\beta _{z}^{\prime }=0.012103$
(or $\gamma _{z}=121$ ms$^{-2}$), at $z=30$, $\beta _{z}^{\prime }=0.00191447$
(or $\gamma _{z}=19.15$ ms$^{-2}$), at $z=100$, $\beta _{z}^{\prime}=0.000191857$ (or $\gamma _{z}=1.92$ ms$^{-2}$),
at $z=200$, $\beta_{z}^{\prime }=4.89903\times 10^{-5}$ (or $\gamma _{z}=5\times 10^{-1}$ ms$^{-2}$),
at $z=625$, $\beta _{z}^{\prime }=5.08713\times 10^{-6}$ (or $\gamma_{z}=5.1\times 10^{-2}ms^{-2}$), at $z=740$, $\beta _{z}^{\prime}=3.6325\times 10^{-6}$
(or $\gamma _{z}=3.6\times 10^{-2}$ ms$^{-2}$), at $z=4\times 10^{3}$, $\beta _{z}^{\prime }=1.24875\times 10^{-7}$
(or $\gamma_{z}=1.25\times 10^{-3}$ ms$^{-2}$), at $z=10^{4}$,
$\beta _{z}^{\prime}=1.992\times 10^{-8}$ (or $\gamma _{z}=2\times 10^{-4}$ ms$^{-2}$), at $z=3\times 10^{5}$, $\beta _{z}^{\prime }=2.22219\times 10^{-11}$
(or $\gamma_{z}=2.2\times 10^{-7}$ ms$^{-2}$), and at $z=10^{6}$, $\beta _{z}^{\prime}=1.999\times 10^{-12}$ (or $\gamma _{z}=2\times 10^{-8}$ ms$^{-2}$).
Finally, we find that $\beta _{z}^{\prime }\rightarrow 0$,  when $z\rightarrow \infty$.

{\bf(ii)} For $\beta _{\rho }^{\prime }$. We see that it begins with
negative values at $z_{i}=-0.25$, $\beta _{\rho i}^{\prime }=-0.0184435$
(or $\gamma _{\rho i}=-184.4$ ms$^{-2}$), then increases until its first
(secondary) maximum, always negative, $\beta _{\rho }^{\prime }=-0.0128256$
(or $\gamma _{\rho }=-128.2$ ms$^{-2}$) at $z=-0.1255$, then decreases until
its minimum $\beta _{\rho }^{\prime }=-0.02304375$ (or $\gamma _{\rho
}=-230.44$ ms$^{-2}$) reached at $z=0.870$. Then it increases again, at $z=2$, $\beta _{\rho }^{\prime }=-0.0160365$ (or $\gamma _{\rho }=-160.4$ ms$^{-2}$),
at $z=4.2579615$, $\beta _{\rho }^{\prime }=0$ (or $\gamma _{\rho }=0$ ms$^{-2}$).
Then it becomes positive, always increasing, until its (main) maximum $\beta _{\rho }^{\prime }=+0.00611618$ (or $\gamma _{\rho }=61.2$ ms$^{-2}$)
reached at $z=8.5275$, from which it decreases again, remaining positive,
towards zero. At $z=10$, $\beta _{\rho }^{\prime }=0.00590862$ (or $\gamma_{\rho }=59.1$ ms$^{-2}$), at $z=30$, $\beta _{\rho }^{\prime }=0.00166611$ (or
$\gamma _{\rho }=16.7$ ms$^{-2}$), at $z=100$, $\beta _{\rho }^{\prime}=1.95353\times 10^{-4}$ (or $\gamma _{\rho }=2$ ms$^{-2}$), at $z=200$,
$\beta_{\rho }^{\prime }=5.14424\times 10^{-5}$ (or $\gamma _{\rho }=0.5$ ms$^{-2}$),
at $z=625$, $\beta _{\rho }^{\prime }=5.45155\times 10^{-6}$ (or $\gamma_{\rho }=5.4\times 10^{-2}$ ms$^{-2}$),
at $z=740$, $\beta _{\rho }^{\prime}=3.89845\times 10^{-6}$ (or $\gamma _{\rho }
=3.4\times 10^{-2}$ ms$^{-2}$), at $z=4\times 10^{3}$, $\beta _{\rho }^{\prime }=1.34888\times 10^{-7}$ (or $\gamma _{\rho }=1.35\times 10^{-3}$ ms$^{-2}$), at $z=10^{4}$,
$\beta_{\rho}^{\prime }=2.16141\times 10^{-8}$ (or $\gamma _{\rho }=2.16\times 10^{-4}ms^{-2}$),
at $z=3\times 10^{5}$, $\beta _{\rho }^{\prime}=2.40387\times 10^{-11}$ (or $\gamma _{\rho }=2.4\times 10^{-7}$ ms$^{-2}$),
at $z=10^{6}$, $\beta _{\rho }^{\prime }=2.16365\times 10^{-12}$
(or $\gamma_{\rho }=2.16\times 10^{-8}$ ms$^{-2}$), and finally $\beta _{\rho }^{\prime}\rightarrow 0$ (or $\gamma _{\rho }=0$ ms$^{-2}$) when $z\rightarrow \infty$.

The two components of the acceleration are positive from $z=4.25$ until the
infinity, with  fairly large positive values until $z\simeq 20$
to $30$ (at $z=10$, $\beta _{z}^{\prime }=0.012$, $\beta _{\rho }^{\prime}=0.006$).

\begin{widetext}

$$%
\begin{tabular}{llllll}
$z$ & $z$ (pc) & $\overset{\cdot \cdot }{z}$ & $\overset{\cdot \cdot }{\rho }$
& $\beta _{z}^{\prime }$ & $\beta _{\rho }^{\prime }$ \\
$-0.25$ & $-0.75\times 10^{-4}$ & $1.57549\times 10^{12}$ & $-8.60382\times
10^{12}$ & $1.68428\times 10^{-2}$ & $-1.84435\times 10^{-2}$ \\
$0.75$ & $\allowbreak 2.\,\allowbreak 25\times 10^{-4}$ &  & $-4.26388\times
10^{11}$ & $6.5\times 10^{-2}$ & $-2.29007\times 10^{-2}$ \\
$0.9$ & $\allowbreak 2.\,\allowbreak 7\times 10^{-4}$ & $\allowbreak $ &  & $%
6.36627\times 10^{-2}$ & $-2.30357\times 10^{-2}$ \\
$2$ & $6\times 10^{-4}$ & $9\times 10^{9}$ & $-1.103\times 10^{11}$ &  & $%
-1.60365\times 10^{-2}$ \\
$4.25$ & $\allowbreak 1$.$2\,\allowbreak 397\times 10^{-3}$ &  & $%
-7.79134\times 10^{9}$ & $3.28446\times 10^{-2}$ & $0$ \\
$4.95$ & $\allowbreak 1.\,\allowbreak 444\times 10^{-3}$ & $2.71515\times
10^{8}$ & $9.89455\times 10^{7}$ & $2.85532\times 10^{-2}$ & $2.47246\times
10^{-3}$ \\
$10$ & $3\times 10^{-3}$ & $5.98106\times 10^{6}$ & $8.19\times 10^{9}$ & $%
1.2103\times 10^{-2}$ & $5.90862\times 10^{-3}$ \\
$30$ & $9\times 10^{-3}$ & $5272.53$ & $1.89417\times 10^{9}$ & $%
1.91447\times 10^{-3}$ & $1.66611\times 10^{-3}$ \\
$10^{2}$ & $0.03$ & $1.39406$ & $2.03259\times 10^{8}$ & $1.91857\times
10^{-4}$ & $1.95353\times 10^{-4}$ \\
$200$ & $0.06$ & $10^{-2}$ & $5.24775\times 10^{7}$ & $4.89903\times 10^{-5}$
& $5.14424\times 10^{-5}$ \\
$500$ & $0.15$ & $10^{-5}$ & $8.552115\times 10^{6}$ & $7.93574\times 10^{-6}
$ & $8.4861\times 10^{-6}$ \\
$625$ & $\allowbreak 0.18231$ & $0$ & $5.48651\times 10^{6}$ & $%
5.08713\times 10^{-6}$ & $5.45155\times 10^{-6}$ \\
$740$ & $\allowbreak 0.222\,$ & $-1.4\times 10^{-6}$ & $3.91956\times 10^{6}$
& $3.6325\times 10^{-6}$ & $3.89845\times 10^{-6}$ \\
$10^{3}$ & $0.3$ & $-1.1\times 10^{-6}$ & $8.15083\times 10^{6}$ & $%
1.99198\times 10^{-6}$ & $2.14225\times 10^{-6}$ \\
$4\times 10^{3}$ & $1.2$ & $-0.6\times 10^{-7}$ & $1.35023\times 10^{5}$ & $%
1.24875\times 10^{-7}$ & $1.34888\times 10^{-7}$ \\
$10^{4}$ & $3$ & $-0.2\times 10^{-7}$ & $2.16227\times 10^{4}$ & $%
1.992\times 10^{-8}$ & $2.16141\times 10^{-8}$ \\
$3\times 10^{5}$ & $90$ & $-10^{-11}$ & $20$ & $2.22219\times 10^{-11}$ & $%
2.40387\times 10^{-11}$ \\
$10^{6}$ & $300$ & $-1.00002\times 10^{-12}$ & $2.16366$ & $1.999\times
10^{-12}$ & $2.16365\times 10^{-12}$%
\end{tabular}%
$$
\textbf{Table 2.} Some indicative values of proper and observable
acceleration components along the same geodesic than in table 1. See the corresponding plots figures 3 to 6.

\end{widetext}

\section{Discussion and conclusion}

In review of what were presented in the above sections, we find the following:  at the initial, the
observable velocities are NR (very weakly relativistic), and the energy is dominated by the
potential, as can be seen from the high initial values of the
proper acceleration. At the end, when $z\rightarrow \infty $, $\ddot{z}\rightarrow 0$ and $\ddot{\rho }
\rightarrow 0$ (particles are free of gravitation) and $\beta _{z}\rightarrow \beta \equiv [(\gamma^2-1)/\gamma^2]^{1/2}$, $\beta _{\rho }\rightarrow 0$ (UR particles), i.e., at the
infinity, energy became fully kinetic ($E=\gamma$, where $\gamma$ is the Lorentz factor.).
The observable, $z$-acceleration, is positive  all
along the geodesic until the infinity. The $z$-acceleration reaches a
maximum and then decreases until zero. The observable $\rho $-acceleration is negative (attractive force) from the initial  until
$z\approx 4.26$, at which it vanishes, and then becomes positive (repulsive force)
until it reaches its maximum value at $z\simeq 8.5275$ before to tend to zero again.
There is a permanent repulsive force along the two axes from
$z\gtrsim 4.5$  until $z\simeq 20$, and then   tending to zero
at the infinity.

The transition from NR ($\beta \approx 0.5$) to UR ($\beta \approx 0.97$)
is in a sub-parsec scale, in accordance with the recent observations \cite{Lee}, though our model predicts a more abrupt transition closer to the
core. This discrepancy can be understood as follows.
The results of \cite{Lee} are the averages over more than 100 cases,  and it could be
possible that the case of a more powerful jet is different from these averaged ones. However,  for M87
there is no improvement (See section 4.3 of \cite{Asada2}) by suggesting a jet
acceleration from $0.01c$ to $0.97c$ on the $z$-range [$4\times 10^{2}$,$10^{6}$].
There were only two exceptions, observed at $43GHz$ with VLBA
\cite{Acciari,Walker}, indicating possible ultra-relativistic bulk speeds
of the M87 jet from $4\times 10^{-2}$ pc, which is in favor of our results. An
interesting discussion on this point is made by Asada et al. given  in section 4.3
of \cite{Asada2}. Very recently, Mertens et al. \cite{Mertens} highlighted a
triple filamentary structure of the jet at the scales $z\in \lbrack
10^{2},10^{3}]$ with a fast interior stream with $\beta \sim 0.92$. But,
in more general cases, upstream (nearer the origin) results are missing. Lee et al.
noted in their discussions (see section 4 of \cite{Lee}) that there are only 3 data in the
range $10^{-2}$ to $10^{-1}$ pc. Further observations will permit to see if
the hypothesis of a spine component, narrower around the axis and more
energetic, is verified.

We first note that with the high value $10^{6}$  for $E$ we could
explain the early acceleration. However, we also made the same evaluations
of the Lorentz factor $\gamma $ for various values ($20$, $10^{2}$, $10^{3}$
and $10^{9}$) of the energy $E$. The results, which  we did not reproduce here, actually remain the same, and in particular
 the transition zones (from NR to UR)
and the slopes do not change dramatically. 
A general explanation of this discrepancy could be that our calculations are
made in vacuum, while one expects that the jet is inside a magnetized
medium, which produces a braking effect. The main theoretical difficulty is that
there is no known exact Kerr solution inside a medium, so all proposed
solutions, including the magnetic models, remain approximated. However, we
cannot ignore the gravitational effect highlighted here which seems to be
the main one, in the sense that it happens earlier and more abruptly
than the magnetic effect which can transform the  internal energy, including the magnetic energy, into kinetic
energy.

In support of the above explanation, we note that the position $r$
of the origin of the synchrotron emission at a given frequency is evaluated
from the synchrotron power (equation (7) in \cite{Lee}),  which depends on the product $(B\gamma )^{2}$. Thus, for each given altitude $z$, the same measure of the
power $L_{S}$ can correspond to different values of $\gamma$ and $B$, for example  a large value of $\gamma $ and a small value of $B$, as long as
the product $B\gamma$ remains the same. In particular, this implies  that in our model the magnetic field can be smaller than that in
their models. This is consistent with our model in which the magnetic field is
 disruptive as least as possible,  so that the main effect will be gravitational.

The remarkable straightness of the observed jet all along its length ($\approx 10^{3}$ kpc) is ensured by the existence of a  positive
 acceleration at the very early time (from $z\approx 4.5$) on a very large scale
(theoretically until infinity). Though progressively decreasing, the
repulsive acceleration is present all the way to infinity. Besides, the bremsstrahlung
radiation (in its own electromagnetic field) is negligible in our model. In
fact we expect that the gravito-magnetic field \cite{Tsoubelis,Bonnor,Bonnor1,Herrera,Bini,Kraniotis,Chicone,Chicone1,Poirier} plays the role of
a magnetic field for the acceleration and collimation of the jet.

As the Blandford-Payne \cite{Blandford1} mechanism can complement the Blandford-Znajek
effect \cite{Blandford} in the magnetic paradigm for the jet formation, the
Poirier-Matthews mechanism \cite{Poirier}  can complement
our BH effect \cite{Gariel} in the gravitomagnetic paradigm. This explanation of the
formation of extragalactic jets can be tested in the future by  probing the
gravitomagnetic structure inherent to Einstein's general relativity.

In addition, the electromagnetic field plays no role for neutral particles, while a gravitational model, similar to ours, can explain the presence of neutral particles in
the formation of jets. In particular,  this gravitational model can  be useful to model the ejection of VHE particles \cite{Abramowski}, including neutral particles such as neutrinos.

\section*{Acknowledgements}

 Part of the work was    carried out when A.W. was visiting the State University of Rio de Janeiro (UERJ), Brazil. A.W. would like to express his gratitude to UERJ for hospitality.
This work is supported in part by Ci\^encia Sem Fronteiras, No. 004/2013 - DRI/CAPES, Brazil (A.W.); Chinese NSF Grant Nos. 11375153 (A.W.) and 11675145 (A.W.).

\end{document}